\documentclass[aps,showpacs,preprintnumbers,amsmath,amssymb,prb,twocolumn
]{revtex4-1}
\usepackage{txfonts}
\usepackage{graphicx}
\usepackage{dcolumn}
\usepackage{bm}
\usepackage{color}
\def\comment#1{}

\def\slashchar#1{\setbox0=\hbox{$#1$}           
   \dimen0=\wd0                                 
   \setbox1=\hbox{/} \dimen1=\wd1               
   \ifdim\dimen0>\dimen1                        
      \rlap{\hbox to \dimen0{\hfil/\hfil}}      
      #1                                        
   \else                                        
      \rlap{\hbox to \dimen1{\hfil$#1$\hfil}}   
      /                                         
   \fi}                                         %

\def\sigmab{{\mbox{\boldmath $\sigma$}}}
\def\nablab{{\mbox{\boldmath $\nabla$}}}

\begin{document}

\title{Semi-metal-insulator transition on the surface of a topological insulator with
in-plane magnetization}

\author{Flavio S. Nogueira}
\affiliation{Institut f{\"u}r Theoretische Physik III, Ruhr-Universit\"at Bochum,
Universit\"atsstra\ss e 150, DE-44801 Bochum, Germany}

\author{Ilya Eremin}
\affiliation{Institut f{\"u}r Theoretische Physik III, Ruhr-Universit\"at Bochum,
Universit\"atsstra\ss e 150, DE-44801 Bochum, Germany}

\date{Received \today}

\begin{abstract}
A thin film of ferromagnetically ordered material proximate to the surface of a three-dimensional topological
insulator explicitly breaks the time-reversal symmetry of the surface states. For an out-of-plane ferromagnetic order
parameter on the surface, the parity is also broken, since the Dirac fermions become massive. This leads
in turn to the generation of a topological Chern-Simons term by quantum fluctuations.
On the other hand, for an in-plane magnetization the surface states remain gapless for the non-interacting Dirac fermions.
In this work we study the possibility of spontaneous breaking of parity due to a
dynamical gap generation on the surface in the presence of a local, Hubbard-like, interaction of strength $g$ between the Dirac fermions.
A gap and a Chern-Simons term are generated for $g$ larger than some critical value, $g_c$, provided the number of Dirac fermions, $N$, is odd. For
an even number of Dirac fermions the masses are generated in pairs having opposite signs, and no Chern-Simons term
is generated. We discuss our results in the context of recent experiments in EuS/Bi$_2$Se$_3$ heterostructures.
\end{abstract}

\maketitle

\section{Introduction}

Due to its unique properties, topological insulators (TI) \cite{Hasan-Kane-RMP,Zhang-RMP-2011}
are likely to play a major role as a component material in different types of heterostructures.
For instance, with a view towards spintronics applications,\cite{MacDonald-NatMat-review} heterostructures involving ferromagnetic
(FM) materials or magnetic impurities
have been studied both theoretically
\cite{Qi-2008,Nagaosa-2010,Nagaosa-2010-1,Franz-2010,Rosenberg-2010,Belzig-2012,Loss-PRL-2012,Cortijo-2012,Nogueira-Eremin-2012,Qi-2012}
and experimentally.\cite{Hor,Wray,Vobornik,Rader-2012,Checkelsky,Moodera-2012,Moodera} Underlying the many applications of magnetic
heterostructures involving TIs is the so-called axion electrodynamics,\cite{Axions} which was shown to distinguish the
electromagnetic response of TIs from ordinary insulators in an essential way.\cite{Qi-2008} Quite generally, it was shown in
Ref. \onlinecite{Qi-2008} that the Lagrangian describing the electromagnetic response of {\it all}
three-dimensional insulators is given by,
\begin{equation}
\label{Eq:L-EM-TI}
 {\cal L}_{\rm EM}=\frac{1}{8\pi}\left(\epsilon{\bf E}^2-\frac{1}{\mu}{\bf B}^2\right)
+\frac{\alpha}{4\pi^2}\theta~{\bf E}\cdot{\bf B},
\end{equation}
where $\theta$ is in general a scalar field, the so-called axion,\cite{Axions} and $\alpha=e^2/(\hbar c)$ is the
fine-structure constant. In ordinary insulators $\theta$ vanishes, but this is not the case in TIs.\cite{Qi-2008} In its
simplest variant the axion field is uniform and assumes the value $\theta=\pi$ for a bulk time-reversal (TR) invariant TI.\cite{Qi-2008} For
uniform $\theta$ the axion term becomes a surface term, leaving therefore the Maxwell equations unaffected.\cite{Axions}
Despite being a surface term when $\theta$ is uniform, the axion term still plays an important role in finite samples.
Indeed, if we imagine a semi-infinite TI sample extending over $z=-\infty$ up to the surface $z=0$, we can use a
covariant formalism to obtain,
\begin{eqnarray}
S_{\rm axion}&=&\frac{\alpha\theta}{32\pi^2}\int d^4x~\epsilon_{abcd}F^{ab}F^{cd}
\nonumber\\
&=&\frac{\alpha\theta}{32\pi^2}\int d^4x~\partial^a(\epsilon_{abcd}A^b\partial^c A^d),
\end{eqnarray}
with the Latin indices running over four-dimensional spacetime. Application of Gauss theorem yields,
\begin{equation}
\label{Eq:Axion-CS}
 S_{\rm axion}=\frac{\alpha\theta}{8\pi^2}\int d^3x~\epsilon_{\mu\nu\lambda}A^\mu\partial^\nu A^\lambda,
\end{equation}
where the Greek indices run over $0$, $x$, and $y$. The above axion action at the surface actually represents a Chern-Simons (CS)
term.\cite{CS} As the Chern-Simons term does not depend on the metric, {\it i.e.} on the geometry of the sample, its presence
can be considered as a manifestation of the topological insulator.

When some symmetry breaking is induced on the topological surface, the axion term may cause significant modifications on
the dynamics of order parameters. For example, if the TI is in contact with a FM material and a proximity-induced magnetization
arises on the topological surface, the magnetization dynamics is modified
\cite{Franz-2010,Rosenberg-2010,Loss-PRL-2012,Nogueira-Eremin-2012}
due to the so-called topological
magnetoelectric (TME) effect,\cite{Qi-2008} consisting of an electric field-induced magnetization caused by the
quantum spin-Hall effect. Although the axion term with $\theta$ uniform does not modify the Maxwell equation, it does modify
the Landau-Lifshitz equation for the magnetization precession on the topological surface.
\cite{Franz-2010,Rosenberg-2010,Loss-PRL-2012,Nogueira-Eremin-2012}

There are also situations where a non-uniform $\theta$ is relevant, like for
example in the case of magnetic fluctuations coupled to the electromagnetic field.\cite{Dyn-Axion} Another
example is when two topological surfaces of the material
are gapped and an external magnetic field induces multichannel edge states.\cite{Rosch-Fritz-2012}
Also in effective theories of topological superconductors
a dynamical axion field plays an important role.\cite{Qi-Witten-Zhang} In all these cases the Maxwell equations are
modified as well and in addition a dynamical field equation for the axion arises.

In order to generate an electromagnetic response featuring an axion term, the helical states have to gap. This may be achieved
by an out-of-plane exchange field which may be induced by proximity effect. This means that the Dirac fermions
on the TI surface become massive and
integrating them out generates a CS term, Eq. (\ref{Eq:Axion-CS}), having $\theta=\pi$.\cite{Nogueira-Eremin-2012} Thus,
in this case TR and parity symmetries are broken on the TI surface, but are still preserved in the bulk.\cite{Qi-2008}
To understand why the mass term breaks the TR and parity symmetries,  observe that the QED-like theory
emerging from the proximity-induced ferromagnetism on the surface of three-dimensional TI (see Sect. II) 
features two-component Dirac fermions and, for this reason, does not have a
chiral symmetry, since $\gamma^5$-like matrices can only be defined for representations featuring
four-component spinors.\cite{ZJ} Indeed, for $2\times 2$ $\gamma$-matrices it is not possible to find 
an additional matrix that anticommutes with
all of them. 
Hence, the massless case corresponding to the case of in-plane magnetization
has no {\it internal} symmetry that would prevent the 
addition of a mass term.  On the other hand, a mass term breaks discrete spacetime symmetries. This case 
corresponds to an out-of-plane magnetization, which indeed is associated to mass term that   
breaks parity and TR symmetries. 
In particular, in 2+1 dimensions parity is realized in terms of a
reflection (mirror symmetry), for example, $x=(x_0,x_1,x_2)\to (x_0,-x_1,x_2)$. Note that inversion of both $x_1$ and $x_2$ does not work,
since this is equivalent to a rotation by $\pi$. In this case the Dirac fermions transform under parity like $\psi\to\gamma^1\psi$,
$\bar \psi\to -\bar \psi \gamma^1$, and $\bar \psi\psi\to -\bar \psi\psi$. The mass term is therefore not invariant under
parity.\cite{Appelquist-1986,Semmenoff} In addition, the TR symmetry, defined by
$\psi\to\gamma^2\psi$, $\bar \psi\to -\bar \psi\gamma^2$,\cite{Appelquist-1986}  is also broken once the mass term is introduced.
This breaking of parity and TR symmetries by massive two-component Dirac fermions causes 
a Chern-Simons (CS) term \cite{CS} to be generated upon integrating them out. 
This generation of a CS term is related to the TME effect if
these Dirac fermions in 2+1 dimensions are viewed as surface states of a three-dimensional
TI.\cite{Qi-2008} It has been shown recently 
that the generation of the CS term by fermionic quantum fluctuations significantly affects
the magnetization dynamics.\cite{Nogueira-Eremin-2012}

However, when only in-plane exchange is present, the Dirac fermions on the TI surface remain massless, thus not violating
TR or parity. Consequently, in this case a CS term is not expected to be generated on the TI surface. An interesting question to be asked
is whether masses for the Dirac fermions simultaneously with the CS term can be spontaneously generated by some symmetry breaking mechanism.
We recall that there are several examples of {\it dynamical} mass generation in QED in 2+1 dimensions \cite{Pisarski84,Appelq}
and related theories, including
some condensed matter models for graphene \cite{Khveschchenko01,Gusynin06,Drut,Muramatsu}
where the Coulomb interaction is taken
into account,\cite{interac-graphene,Sheehy-2007,DTSon,Herbut-PRB-2009} and theories for the pseudogap in high-$T_c$ (cuprate)
superconductors.\cite{Kim,Rantner,Zlatko-CSB,Herbut-AFL,FT} However, the latter theories feature an even number of Dirac cones, allowing the
use of four-component Dirac spinors. Therefore, they have a chiral symmetry,\cite{Appelq} since in this case
two $\gamma^5$-like matrices anticommuting with all
$\gamma$-matrices can be defined (see Sect. \ref{Sec:Seff-O-o-P}), and this is simply not 
possible with an odd number of Dirac cones arising
in TIs.\cite{Hasan-Kane-RMP,Zhang-RMP-2011}
Thus, in none of the mentioned models a CS term can be generated when masses for the Dirac fermions
are dynamically generated.

In this paper we analyze what happens for an interacting TI having an odd number of Dirac fermions
in the proximity to a FM inducing an in-plane exchange. In particular, we show that in the
presence of a screened Coulomb interaction, a mass for the Dirac fermions is dynamically generated
only if the interaction strength exceeds some critical value.
Under the same conditions a CS term is also generated.
As a result, the dynamical generation of the mass due to screened Coulomb interaction in the case
of TI in proximity to in-plane FM yields a TME effect similar to the case of an out-of-plane magnetization.
In agreement with earlier calculations in the context of QED,\cite{Appelq-parity} we also
show that for an even number of Dirac fermions
there is mass generation, but parity and TR are overall preserved and no CS term arises.

The plan of the paper is as follows. In Sect. II we define the QED-like model used in this paper and discuss
its effective action in Sects. \ref{Sec:Seff-O-o-P} and
\ref{Sect:Seff-i-p}. Sect. \ref{Sec:dynmass} contains the main results, i.e., the solution of the gap equation,
showing that a semi-metal insulator transition occurs for a large enough value of the coupling constant. Sect. \ref{Sect:Discussion}
discuses the relation of our results to recent experiments and in Sect. \ref{Sect:Conclusion}
we present the conclusions of this work.
Three appendices contain additional technical information about the calculations.

\section{Model}

\begin{figure}
 \includegraphics[width=9cm]{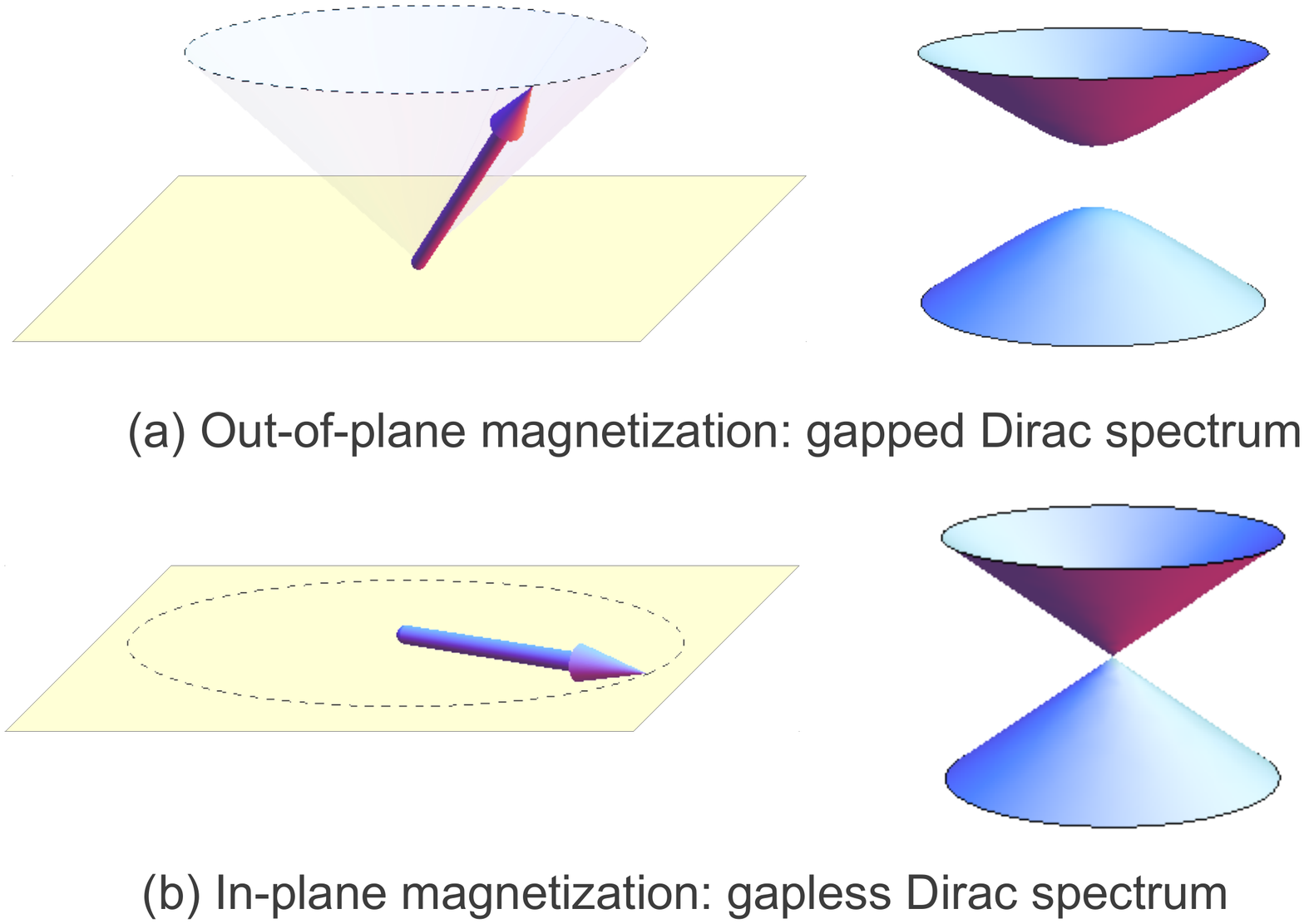}
 \caption{(Color online) Schematic comparison between two types of magnetization orientation on the surface of a TI. (a) In the case of out-plane
 magnetization the electronic spectrum at the surface is gapped. (b) For in-plane magnetization the spectrum is gapless.}
 \label{Fig:precession}
\end{figure}

In first-quantized form the Hamiltonian for a topological surface with strong spin-orbit coupling
in contact with a thin FM layer can be written in a form including a Rashba-like term and an
anisotropic exchange energy,\cite{Loss-PRL-2012}
\begin{equation}
 H=v_F(-i\hbar\nablab\times\hat {\bf z})\cdot{\sigmab}-J(n_x\sigma_x+n_y\sigma_y)-J_\perp n_z\sigma_z,
\end{equation}
where $v_F$ is the Fermi velocity, $\nablab=(\partial_x,\partial_y)$,
and $J$ and $J_\perp$ are the in-plane and out-of-plane exchange energies coupling to the
magnetization ${\bf n}$, respectively.  For an uniform magnetization, the Hamiltonian is easily diagonalized, yielding
the generally gapped energy spectrum,
\begin{equation}
 E_\pm=\pm \sqrt{(p_x-Jn_y)^2+(p_y+Jn_x)^2+J_\perp^2n_z^2},
\end{equation}
where ${\bf p}=\hbar v_F{\bf k}$.
For a vanishing out-of-plane exchange we have a gapless spectrum with a Dirac point at $(Jn_y,-Jn_x)$.
Thus, while for an out-of-plane magnetization the Dirac spectrum is gapped, it is gapless in the case of
in-plane exchange; see Fig. \ref{Fig:precession}.

In order to see whether for  $J_\perp=0$ a mass can be dynamically generated, we have to consider the quantum fluctuations of
the magnetization on the TI surface and, in addition, the Coulomb interaction.
If $\psi=[\psi_\uparrow, \psi_\downarrow]^T$, the Schr\"odinger equation, $i\hbar\partial_t\psi=H\psi$, for
a vanishing out-of-plane exchange in the absence of Coulomb interaction reads,
\begin{equation}
 \sigma_zi\hbar\partial_t\psi=\sigmab\cdot(v_F\hbar\nablab-iJ{\bf a})\psi,
\end{equation}
where ${\bf a}=(n_y,-n_x)$ plays the role of a vector potential.
The above Schr\"odinger equation has actually the form of a
Dirac equation in the presence of an electromagnetic field. Thus, the Lagrangian of the TI surface proximate to
a FM thin film inducing a planar magnetization on it is given by,
\begin{equation}
\label{Eq:TI-QED}
 {\cal L}_0=\bar \psi[i\gamma_0\hbar\partial_t-i\vec \gamma\cdot(v_F\hbar\nablab+iJ{\bf a})]\psi,
\end{equation}
 where
$\gamma^0=\sigma_z$, $\gamma^1=-i\sigma_x$, and $\gamma^2=i\sigma_y$. The above Lagrangian has
a QED-like form in $d=2+1$ dimension with a vector potential ${\bf a}=(n_y,-n_x)$, and no time component for
the gauge field. A time component for the gauge field is introduced if we assume a screened Coulomb interaction on the TI surface with
interaction Hamiltonian density
\begin{equation}
 {\cal H}_{\rm int}=\frac{g}{2}(\psi^\dagger\psi)^2=\frac{g}{2}(\bar \psi\gamma^0\psi)^2,
\end{equation}
where $g>0$ and $\bar \psi=\psi^\dagger\gamma^0$ as usual. Then, the full Lagrangian acquires the following form
\begin{equation}
\label{Eq:L}
 {\cal L}={\cal L}_0-{\cal H}_{\rm int},
\end{equation}
which can be rewritten in terms of an auxiliary field, $a_0$, via a Hubbard-Stratonovich (HS) transformation to obtain,
\begin{equation}
 \label{Eq:TI-QED-1}
 {\cal L}=\bar \psi(i\slashchar{\partial}-J\slashchar{a})\psi-\frac{J^2}{2g}a_0^2,
\end{equation}
where we have used the standard Dirac slash notation, $\slashchar{Q}=\gamma^\mu Q_\mu$.
Note that the vector field $a_\mu$ is not dynamical at
the Lagrangian level, since the only quadratic term in the gauge field $a_\mu$ is a term proportional to $a_0^2$ with
a time-independent coefficient.
This term implies that the gauge symmetry is broken in the temporal direction.

We are disregarding the long-range contribution to the Coulomb interaction because
it has been shown to be irrelevant in
the long wavelength limit in theoretical studies of
interacting graphene \cite{interac-graphene,Sheehy-2007,Herbut-PRB-2009} and a similar reasoning also applies here.
From a model point of view, our Lagrangian corresponds to a restricted Thirring model,\cite{Thirring}
in the sense that only the zeroth component of the current $j_\mu=\bar \psi\gamma_\mu\psi$
appears squared in the interaction.

One important consequence of the term quadratic in $a_0$, is that $g$ does
not renormalize. This follows from the gauge symmetry of the fermionic sector and can be easily proved using Ward identities.\cite{ZJ}
An easy way of seeing this without making explicit use of the Ward identity is to introduce renormalized fields $\psi_r=Z^{-1/2}\psi$ and
$a^\mu_r=Z_a^{-1/2}a^\mu$ and observe that gauge invariance of the fermionic sector implies that the renormalized exchange coupling is
$J_r=\sqrt{Z_a}J$, otherwise the form of the covariant derivative would not be preserved by a gauge transformation.
Furthermore, current conservation implies that any fluctuation correction for $a^\mu$ is transverse, and therefore
terms quadratic in $a^\mu$ which are not gauge invariant do not renormalize, yielding
$J_r^2/g_r=Z_aJ^2/g$ and consequently $g_r=g$. Therefore, $g$ is a good tuning parameter in our theory. The
fact that $g$ does not renormalize will be important in our subsequent analysis.

Note that our Lagrangian does not include an intrinsic dynamics for the magnetization. Although the FM above the TI surface has
its own dynamics, we are assuming a minimal model on the topological surface where the only exchange interaction is the
one between the electronic spin and the surface magnetization. Thus, the whole magnetization dynamics on the topological
surface will be generated by the quantum fluctuations of the Dirac fermions. It is certainly important to
include other exchange effects, like it was done in Refs. \onlinecite{Nagaosa-2010} and \onlinecite{Nogueira-Eremin-2012}.
However, our main aim here is to study the dynamical mass generation and the spontaneous breaking of parity and TR symmetries.
For this purpose our minimal exchange model (\ref{Eq:TI-QED-1}) already exhibits this feature and has
the advantage of being analytically more tractable.

\section{Effective action in the presence of out-of-plane exchange}
\label{Sec:Seff-O-o-P}

\subsection{Effective theory}

Let us first recall the situation for $J_\perp\neq 0$ that is when the ferromagnet has the out-plane component of the magnetization.
Since we are assuming that the FM above the TI surface is in the broken symmetry state, we can write
${\bf n}=\langle n_z\rangle\hat {\bf z}+{\bf n}_\perp$, where $\langle n_z\rangle\neq 0$. which can be either positive or negative, and
${\bf n}_\perp=(n_x,n_y)$ are small transverse fluctuations.
In this case the Lagrangian (\ref{Eq:TI-QED-1}) becomes
\begin{equation}
 \label{Eq:TI-QED-2}
 {\cal L}=\bar \psi(i\slashchar{\partial}-J\slashchar{a}-m)\psi-\frac{J^2}{2g}a_0^2,
\end{equation}
where $m=J_\perp\langle n_z\rangle$.
As discussed in the Introduction, the mass term  {\it explicitly} breaks parity and time-reversal symmetry.
Let us assume that we have $N$ Dirac fermion species and work within an
imaginary time formalism.
In this case after integrating out the $N$ fermionic degrees of freedom, we obtain,
\begin{equation}
 S_{\rm eff}=-N{\rm Tr}\ln(\slashchar{\partial}-iJ\slashchar{a}+m)+\frac{J^2}{2g}\int d^3x~ a_0^2.
\end{equation}
Now we expand the above effective action up to quadratic order in the vector field $a_\mu$, which
in momentum space reads,
\begin{equation}
 \label{Eq:S-eff-0}
 S_{\rm eff}=\frac{1}{2}\int\frac{d^3p}{(2\pi)^3}\left[\Sigma_{\mu\nu}(p)a_\mu(p)a_\nu(-p)
 +\frac{J^2}{g}a_0(p)a_0(-p)\right],
\end{equation}
where $p=(\omega,v_F{\bf p})$ and $\Sigma_{\mu\nu}(p)$ is the one-loop vacuum polarization, which is evaluated
in detail in Appendix \ref{App:vacpol}. The result is
\begin{eqnarray}
 \Sigma_{\mu\nu}(p)&=&\frac{NJ^2}{2}
 \left[\frac{|m|}{2\pi}+(p^2-4m^2)I(p)\right]\left(\delta_{\mu\nu}-\frac{p_\mu p_\nu}{p^2}\right)
 \nonumber\\
 &-&2NJ^2mI(p)\epsilon_{\mu\nu\lambda}p_\lambda,
\end{eqnarray}
where
\begin{equation}
 I(p)=\frac{1}{4\pi|p|}\arctan\left(\frac{|p|}{2|m|}\right).
\end{equation}
Thus, in the long wavelength regime $|p|\ll |m|$ [see Eq. (\ref{Eq:large-mass}) in Appendix A] the effective action
in real time is given by,
\begin{equation}
 S_{\rm eff}\approx\frac{NJ^2}{8\pi}\int d^3x\left[-\frac{1}{6|m|}f_{\mu\nu}f^{\mu\nu}+\frac{m}{|m|}
 \epsilon_{\mu\nu\lambda}a^\mu\partial^\nu a^\lambda\right],
 \label{CS-term}
\end{equation}
where $f_{\mu\nu}=\partial_\mu a_\nu-\partial_\nu a_\mu$.
In Ref. \onlinecite{Nogueira-Eremin-2012} this result was used for the case where $m>0$.
We note also that the first (Maxwell) term in Eq. (\ref{CS-term}) contains a dimensionfull coefficient as it depends on $|m|$. Thus,
this term is non-universal and receives corrections from all orders in perturbation theory.
The second term (the CS term) is universal and its prefactor  $m/|m|$ is just a sign. Moreover, being
independent of the metric, it is not expected to be modified by the scale
transformations, so it does not renormalize. The Coleman-Hill theorem \cite{Coleman-Hill}
on the non-renormalization of the CS term provides a more precise statement of this argument.

Observe that the number of Dirac fermions, $N$, must be necessarily odd, otherwise no CS term is generated.
Although this point was noticed before in Ref. \onlinecite{Qi-2008} (see Sect. IV-D there),
we would like to revisit it in the framework of
our calculations. In order to see the effect, let us now assume that each of the $N$ Dirac fermions has a mass, $m_i$
($i=1,\dots,N$). It is straightforward to see that the low-energy form of the CS term is now given by,
\begin{equation}
 S_{\rm CS}=\frac{J^2}{8\pi}\left(\sum_{i=1}^N\frac{m_i}{|m_i|}\right)\int d^3x \epsilon_{\mu\nu\lambda}a^\mu\partial^\nu a^\lambda.
\end{equation}
Now, if the number of Dirac fermions is even, we can rewrite the Dirac Lagrangian in terms of $N/2$ four-component
Dirac fermions using $4\times 4$ $\gamma$-matrices. In this case a mass term $m_i\bar \psi\psi$ is invariant
under both parity and time-reversal transformations, just like in the case of QED in four-dimensional spacetime.
Namely, when four-component Dirac fermions are introduced in 2+1 dimensions, it is possible to introduce $4\times 4$ Dirac matrices
of the form,\cite{Appelquist-1986}
\begin{equation}
\gamma^0=\left(
\begin{array}{cc}
\sigma_z & 0\\
\noalign{\medskip}
0 & -\sigma_z
\end{array}
\right),~~~~~~~~~
\gamma^1=\left(
\begin{array}{cc}
i\sigma_x & 0\\
\noalign{\medskip}
0 & -i\sigma_x
\end{array}
\right),
\nonumber
\end{equation}
\begin{equation}
\label{Eq:gamma}
\gamma^2=\left(
\begin{array}{cc}
i\sigma_y & 0\\
\noalign{\medskip}
0 & -i\sigma_y
\end{array}
\right),
\end{equation}
where $\sigma_x$, $\sigma_y$, and $\sigma_z$ are the Pauli matrices.
With the above representation the chiral symmetry can be defined via the following two matrices,\cite{Pisarski84,Appelq}
\begin{equation}
\gamma^3=i\left(
\begin{array}{cc}
0 & I_2\\
\noalign{\medskip}
I_2 & 0
\end{array}
\right),~~~~~~~~~
\gamma^5=i\left(
\begin{array}{cc}
0 & I_2\\
\noalign{\medskip}
-I_2 & 0
\end{array}
\right),
\nonumber
\end{equation}
anticommuting with all $\gamma$-matrices (\ref{Eq:gamma}), where $I_2$ is a $2\times 2$ identity matrix. The Lagrangian for
massless four-component Dirac fermions
has therefore the invariance under the chiral transformations $\psi\to e^{i\theta\gamma^3}\psi$ and $\psi\to e^{i\phi\gamma^5}\psi$.
Since under these transformations $\bar \psi\to \psi^\dagger e^{-i\theta{\gamma^3}^\dagger}\gamma^0=\bar \psi e^{i\theta\gamma^3}$ and
$\bar \psi\to \psi^\dagger e^{-i\phi{\gamma^5}^\dagger}\gamma^0=\bar \psi e^{i\phi\gamma^5}$, the current $j^\mu=\bar \psi\gamma^\mu\psi$
is invariant, but $\bar \psi\psi$ is not. Thus, for massless QED in 2+1 dimensions with four-component Dirac fermions
the chiral symmetry prevents the addition of a mass term.\cite{Pisarski84,Appelq}
Indeed, a term $m\bar \psi\psi$ in the Lagrangian would explicitly break the chiral symmetry, but
not parity and TR.
In our particular case this implies that we must have $\sum_i m_i/|m_i|=0$ in
the representation where $N=2n$ two-component Dirac fermions are used, since this is equivalent to $n$ four-component Dirac
fermions.
Thus, the spectrum must consist of $N/2$ masses having
the opposite sign of the remaining $N/2$ ones. We conclude that $N$ must be odd in order to generate a CS term, and
we can write $N=2n+1$ ($n=0,1,\dots,(N-1)/2$). This is consistent with the fact that a TI features an odd number of Dirac fermions. Therefore,
the CS action originating from the coupling of Dirac fermions to an out-of-plane exchange can be written in the form,
\begin{equation}
 \label{Eq:CS-action}
 S_{\rm CS}=\frac{J^2}{4\pi}\left(n+\frac{1}{2}\right)\frac{m}{|m|}\int d^3x \epsilon_{\mu\nu\lambda}a^\mu\partial^\nu a^\lambda,
\end{equation}
which emphasizes the role of an odd number of Dirac fermions.
Note that the apparent ``quantization'' of the CS coefficient in Eq. (\ref{Eq:CS-action}) does not have in
the present context the same origin as the Hall conductivity in the quantum Hall effect, as
it arises from integrating out Dirac fermions coupled to a vector field. However, it is consistent
with the analysis of Qi {\it et al.}\cite{Qi-2008} of the axion term, since we find here that the axion field has the
constant value $\theta=\pi$. Indeed, for a single Dirac fermion ($n=0$), Eq. (\ref{Eq:CS-action}) coincides
with Eq. (\ref{Eq:Axion-CS}) for this value of $\theta$ and identifying $\alpha=J^2$, so the in-plane exchange coupling
squared plays the role of the fine-structure constant in this case. It is worth to mention in this context yet
another aspect of the problem discussed recently in Ref. \onlinecite{Rosch-Fritz-2012}, namely, the dependence
of the Hall conductivity at the edges of a finite sample for a general value of $\theta$, not necessarily $0$ or $\pi$.
This case is relevant if there is an external magnetic field.
In this situation it can be shown that the Hall conductivity is given by $\sigma_H=e^2/(2\pi)[n+\theta/(2\pi)]$
[in units where $\hbar=1$; here we are assuming that the lowest value of $n$ is zero, just like
in Eq. (\ref{Eq:CS-action})], and quantization will hold if $\theta$ changes by $\pm 2\pi(n+1)$
on loops containing $n+1$ edges channels.\cite{Rosch-Fritz-2012}

It is important to emphasize here the difference between a non-Abelian CS action, where the CS coupling is quantized due
to the requirement of gauge-invariance,\cite{CS} so that the integer
arising there is actually a winding number. Note that the derivation of the
axion term in the bulk leads to an expression for $\theta$ given in terms of a non-Abelian Berry connection in momentum space, which
is derived in terms of the actual band structure on the bulk.\cite{Qi-2008} This is a topological invariant generalizing
the Thouless-Kohmoto-Nightingale-Nijs invariant,\cite{TKNN} which features a Berry Abelian
curvature in momentum space, to higher dimensions. In our calculations done in the 2+1 dimensions $\theta$ is also quantized in the sense that it may
have only two values, $\pi$ or 0, where the latter refers to the absence of the CS term.

\subsection{Magnetization dynamics}

Rewriting the CS term in explicitly in terms of components allows us to analyze its physical content relative to
the magnetization dynamics on the topological surface:\cite{Nogueira-Eremin-2012}
\begin{equation}
 \label{Eq:Berry}
 S_{\rm CS}=\frac{NJ^2 \theta}{8\pi^2}\int dt\int d^2r(n_y\partial_tn_x-n_x\partial_tn_y-2{\bf n}\cdot{\bf E}),
\end{equation}
where ${\bf E}=-\nablab a_0$ yields the electric field associated to the screened Coulomb potential and
${\bf n}=(n_x,n_y,m/J_\perp)$. We observed that the CS action contains an induced Berry phase associated to
the precession of the magnetization.\cite{Nagaosa-2010,Nogueira-Eremin-2012}
If we neglect for a moment the contribution of the
Maxwell term in the effective action, we obtain simply,
\begin{equation}
 \partial_tn_i=\epsilon_{ij}E_j,
\end{equation}
which is the expected result for a spin-Hall response. In order to obtain the full magnetization precession, we have also
to consider the fluctuations in $n_z$ around its expectation value $\langle n_z\rangle$. This was done in Ref.
\onlinecite{Nogueira-Eremin-2012}. The result is a Landau-Lifshitz equation where in addition to
the usual torque $\gamma({\bf n}\times{\bf H}_{\rm eff})$ yielding the precession around
the effective magnetic field ${\bf H}_{\rm eff}$, a magnetoelectric torque $\sim{\bf n}\times{\bf E}$ arises.

\section{Effective action for in-plane exchange}
\label{Sect:Seff-i-p}

When $J_\perp=0$ the Dirac fermions are massless. Thus, in this regime
the effective action becomes,
\begin{eqnarray}
 \label{Eq:S-eff}
 S_{\rm eff}&=&\frac{1}{2}\int\frac{d^3p}{(2\pi)^3}\left[\phantom{\frac{}{}}\Pi(p)\left(p^2\delta_{\mu\nu}-p_\mu p_\nu\right)a_\mu(p)a_\nu(-p)
 \right.\nonumber\\
 &+&\left.\frac{J^2}{g}a_0(p)a_0(-p)\right],
\end{eqnarray}
where $\Pi(p)=NJ^2/(16|p|)$ is the usual vacuum polarization for massless Dirac fermions in 2+1 dimensions.
From Eq. (\ref{Eq:S-eff}) we derive the propagator (see Appendix B),
\begin{eqnarray}
\label{Eq:gauge-prop}
 D_{\mu\nu}(p)&=&\langle a_\mu(p)a_\nu(-p)\rangle=\frac{1}{p^2\Pi(p)}\left\{\phantom{\frac{}{}}\delta_{\mu\nu}
\right.\nonumber\\
&+&\left.\left[\frac{g}{J^2}p^2\Pi(p)+1\right]\frac{p_\mu p_\nu}{\omega^2}
-\frac{(p_\mu\delta_{\nu 0}+p_\nu\delta_{\mu 0})}{\omega}
\right\}.
\end{eqnarray}
Interestingly, this result shows that a vector field with a mass term  only along the temporal direction is not gapped.
With an isotropic mass term of the form $(J^2/g)a_\mu a_\mu$, we would obtain instead
$D_{\mu\nu}(p)=[p^2\Pi(p)+J^2/g]^{-1}[\delta_{\mu\nu}+(g/J^2)\Pi(p)p_\mu p_\nu]$, which is clearly gapped.
As we will see shortly, this difference is important, as in our case two of the components of the vector field relate
to the magnetization and magnetic excitations are supposed to be gapless.

The propagator (\ref{Eq:gauge-prop}) does not smoothly connect to the strongly coupled regime, $g\to\infty$.
This is a typical behavior for massive vector fields \cite{ZJ} which is also reflected here, although our vector field is
only massive along the temporal direction. Note, however, that in the strongly coupled regime our model reduces to a QED model in
2+1 dimensions with two-component Dirac fermions. As mentioned earlier, no CS term is generated in this case when
the Dirac fermions become gapped.\cite{Appelq-parity}

The purely magnetic effective action is finally obtained by
integrating out $a_0$ in the effective action (\ref{Eq:S-eff}). This yields,
\begin{eqnarray}
\label{Eq:S-eff-1}
S_{\rm eff}^{\rm FM}&=&\frac{1}{2}\int\frac{d^3p}{(2\pi)^3}\Pi(p)\left[p^2\delta_{ij}-F(p)v_F^2p_ip_j\right]
a_i(p)a_j(-p)\nonumber\\
&=&\frac{1}{2}\int\frac{d^3p}{(2\pi)^3}\Pi(p)\left\{\left[p^2-F(p)v_F^2{\bf p}^2\right]{\bf n}(p)\cdot{\bf n}(-p)
\right.\nonumber\\
&+&\left.F(p)v_F^2[{\bf p}\cdot{\bf n}(p)][{\bf p}\cdot{\bf n}(-p)]
\right\},
\end{eqnarray}
where
\begin{equation}
 F(p)=\frac{p^2\Pi(p)+J^2/g}{
v_F^2{\bf p}^2\Pi(p)+J^2/g}.
\end{equation}
Going back to real time, the magnetic susceptibility $\chi(\omega,{\bf p})=\langle n_+(\omega,{\bf p})n_-(\omega,{\bf p})\rangle$,
where $n_\pm=n_x\pm n_y$,
is determined from Eq. (\ref{Eq:S-eff-1}) as,
\begin{eqnarray}
 \chi(\omega,{\bf p})&=&\frac{16}{NJ^2\sqrt{v_F^2{\bf p}^2-(\omega+i\delta)^2}}
 \nonumber\\
 &\times&\left\{1
 -\frac{Ngv_F^2{\bf p}^2}{(\omega+i\delta)^2}
 \left[1+\frac{16}{Ng}\frac{\sqrt{v_F^2{\bf p}^2-(\omega+i\delta)^2}}{v_F^2{\bf p}^2}\right]\right\}.
\end{eqnarray}
From the pole of $\chi(\omega,{\bf p})$ we infer that the spin-wave velocity is identical to
the Fermi velocity. This is the consequence of our approximation as we ignored the bare spin dynamics of the ferromagnet at the interface and our spin excitations are itinerant excitations due to Dirac fermions. In this case, we also see that by comparing with the scaling
behavior $\chi(\omega,{\bf p})\sim[v_F^2{\bf p}^2-(\omega+i\delta)^2]^{\eta/2-1}$ for $\omega$ near $v_F|{\bf p}|$, yields an
anomalous scaling dimension $\eta=1$. This induced anomalous dimension on the topological surface is very different from the one of a
two-dimensional planar
FM at $T=0$ corresponding to a three-dimensional ($d=2+1$) XY universality class,
having $\eta\approx 0.04$.

\section{Dynamical generation of out-of-plane exchange and spontaneous breaking of
parity and time-reversal invariance}
\label{Sec:dynmass}

Next, we consider the fermionic propagator.
Within an imaginary time formalism, the fermion propagator $G(p)$ is given in general form by
\begin{equation}
\label{Eq:F-prop}
 G^{-1}(p)=i\gamma_\mu p_\mu+J\int\frac{d^3k}{(2\pi)^3}\gamma_\mu G(p-k)D_{\mu\nu}(k)\Gamma_\nu(p,k),
\end{equation}
where $\Gamma_\nu(p,k)$ is the vertex function. It is understood that the Dirac matrices above are the imaginary time
counterparts of the real time ones defined earlier. They are assumed to satisfy the Clifford
algebra $\gamma_\mu\gamma_\nu+\gamma_\nu\gamma_\mu=2\delta_{\mu\nu}$. In order to determine $G(p)$ approximately, we
make the decomposition $G^{-1}(p)=Z(p)i\gamma_\mu p_\mu+\Sigma(p)$ and assume the lowest order form for the vertex function,
$\Gamma_\mu(p,k)=J\gamma_\mu$. Furthermore, we will set $Z(p)\approx 1$ for $G(p)$ inside the integral in Eq. (\ref{Eq:F-prop}).
A mass will be generated if $m\equiv \Sigma(0)$ does not vanish. Note that a non-vanishing $m$ implies
that $\langle\bar \psi\psi\rangle\neq 0$. Since $\bar \psi\psi=n_\uparrow-n_\downarrow$, mass generation
implies also an emergent third component of the magnetization. Our strategy will be to make an approximation in which $\Sigma(p)$ is
uniform, $\Sigma(p)=\Sigma(0)=m$, and see whether there is a solution to Eq. (\ref{Eq:F-prop}) with $m\neq 0$.
We will solve Eq. (\ref{Eq:F-prop}) under the assumption that $m\ll|p|\ll\Lambda$, where $\Lambda$ is an
ultraviolet cutoff, which here is naturally given by $\Lambda\approx NJ^2/(\hbar v_F^2)$, similarly to QED in 2+1 dimensions,\cite{Appelquist-1986}
where the cutoff is determined by the charge squared times the number of fermion components.
The fermion mass modifies the vacuum polarization, and now a term odd under parity may arise, so that the photon
self-energy becomes,
\begin{equation}
\Sigma_{\mu\nu}(p)=\left(\delta_{\mu\nu}-\frac{p_\mu p_\nu}{p^2}\right)\Sigma_{\rm even}(p)+\epsilon_{\mu\nu\lambda}p_\lambda
\Sigma_{\rm odd}(p)
\end{equation}
However, under the assumption  $m\ll|p|\ll\Lambda$,
the contribution that is even under parity and time-reversal, corresponding to the transverse term in Eq. (\ref{Eq:S-eff}),
remains unchanged, and we obtain
$\Sigma_{\rm even}(p)=p^2\Pi(p)$.

In order
to investigate the gap equation, we follow Ref. \onlinecite{Appelq-parity} and assume that $N-L$ fermions acquire a positive
mass $+m$, while the remaining $L$ fermions acquire a negative mass $-m$. Thus, the effective action
(\ref{Eq:S-eff}) for the vector field receives the following additional contribution odd under parity and time-reversal
[see Eq. (\ref{Eq:small-mass}) in Appendix A],
\begin{equation}
\label{Eq:Sodd}
 S_{\rm eff}^{\rm odd}=\frac{2}{N}\sum_{i=1}^Nm_i\int\frac{d^3p}{(2\pi)^3}\Pi(p)\epsilon_{\mu\nu\lambda}p_\lambda a_\mu(p)a_\nu(-p).
\end{equation}
This leads in turn to an additional term in the vector field propagator given by
$D_{\mu\nu}^{\rm odd}(p)=-32\sum_i(m_i/N)\epsilon_{\mu\nu\lambda}p_\lambda/(NJ^2|p|^3)$.
Thus, the following self-consistent equation for $m_i$
is obtained,
\begin{eqnarray}
 \label{Eq:self}
1&=&\frac{16}{N}\left\{\left[1-8~\frac{m}{m_i}~\left(\frac{N-2L}{N}\right)\right]\int\frac{d^3k}{(2\pi)^3}\frac{1}{|k|(k^2+m^2)}
\right.\nonumber\\
&+&\left.\int\frac{d^3k}{(2\pi)^3}\frac{|k|}{\omega^2(k^2+m^2)}
\right\}
+g\int\frac{d^3k}{(2\pi)^3}\frac{k^2}{\omega^2(k^2+m^2)}.\nonumber\\
\end{eqnarray}
The second and third integrals above require some care, although they are not difficult to solve, and
are calculated in the Appendix C.

After performing
all integrals, the gap equation for $i=1,\dots, N-L$ becomes,
\begin{equation}
 \label{Eq:gap-1}
 1-\frac{8}{\pi^2N}=
 \frac{64}{\pi^2N}\left(1-\frac{2L}{N}\right)\ln(|m|/\Lambda)+\frac{g\Lambda}{4\pi}\left(1+\frac{1}{\pi}-3\frac{|m|}{\Lambda}\right),
\end{equation}
while for $i=N-L+1,\dots, N$, we obtain,
\begin{equation}
 \label{Eq:gap-2}
 1-\frac{8}{\pi^2N}
 =-\frac{64}{\pi^2N}\left(1-\frac{2L}{N}\right)\ln(|m|/\Lambda)+\frac{g\Lambda}{4\pi}\left(1+\frac{1}{\pi}-3\frac{|m|}{\Lambda}\right).
\end{equation}
To have the solution for $L\neq 0$ one sees that the gap equations (31)-(32) are only compatible with each other if $N$ is even, i.e., $L=N/2$.
For the odd number of fermions $L=0$ which we discuss below.
For an even number of Dirac
fermions we introduce the dimensionless quantities $\hat m=m/\Lambda$ and $\hat g=\Lambda g$ and obtain
\begin{equation}
 \label{Eq:sol-even}
 |\hat m|=\frac{\pi+1}{3\pi}\left(1-\frac{\hat g_c}{\hat g}\right),
\end{equation}
where
\begin{equation}
 \label{Eq:gc}
 \hat g_c=\frac{4\pi^2}{\pi+1}\left(1-\frac{8}{\pi^2N}\right).
\end{equation}
In terms of the dimensionful coupling constant $g$, we can explicitly write
\begin{equation}
\label{Eq:gc-1}
\frac{NJ^2 g_c}{\hbar^2v_F^2}=\hat g_c \quad.
\end{equation}
Note that $g_c$ has dimension of length squared and relates in terms of energy scales of the lattice model via
$g\sim Ua^2/t$, where $a$ is the lattice spacing, and $U$ and $t$ are the Hubbard interaction and hopping, respectively.
In field-theoretic units, $\hbar=v_F=1$, $g$, it has of course dimension of length, since the Dirac fermions
have in this case dimension of (length)$^{-1}$ and the action has to be dimensionless.

Thus, we obtain that if $N$ is even a gap is generated provided $\hat g>\hat g_c$. In this case there is no generation of
CS term. Therefore, parity and time-reversal symmetries remain preserved. The above result distinguishes itself from the QED
case \cite{Appelq-parity} due to a complete cancellation of the logarithm.

For the odd number of Dirac fermions, which is the situation
corresponding to a TI, we now search for the solution $m_i=+m$ for all $i$ and $L=0$. In this case the
logarithmic term survives and dominates for $|m|\ll\Lambda$ over the linear term in $|m|$. Thus, the gap equation is
given by Eq. (\ref{Eq:gap-1}) with $L=0$ and where the term proportional to $|m|/\Lambda$ is neglected.
Then we obtain,
\begin{equation}
\label{Eq:sol-odd}
 \hat m=\exp\left[-\frac{(\pi+1)N}{256}(\hat g-\hat g_c)\right],
\end{equation}
where $\hat g_c$ is the same as before, given by Eq. (\ref{Eq:gc}).
Eq. (\ref{Eq:sol-odd}) only makes sense for $\hat g>\hat g_c$, otherwise it does not decrease with increasing $N$, which would at large $N$
contradict the condition $m\ll|p|\ll\Lambda$, in a situation reminiscent from the QED case.\cite{Appelq-parity} However,
in our case it is possible to overcome the difficulty encountered there and obtain in addition the generation of a CS term.
In other words, we find that the
dynamic generation of the mass due to the screened Coulomb interaction in a TI/FM heterostructure where the FM
has in-plane components has similar consequences for the electrodynamics of the TI in contact to a FM
with out-of-plane magnetization. Namely, the topological magnetoelectric term arises in the former
case when values of the
interaction above $\hat{g_c}$ is reached.
The latter is determined by the bare value of $g$
in the topological insulator, multiplied by the ratio of $NJ^2/(\hbar v_F)^2$; see Eq. (\ref{Eq:gc-1}).
This means that in the TI/FM heterostructure with in-plane
magnetization the dynamic mass generation will be proportional to the absolute value
of the in-plane exchange coupling in the FM. This is interesting as it points towards experimental
realizability of the observed effect by varying the FM substrate of the heterostructure.

\section{Discussion}
\label{Sect:Discussion}

We note that the case of in-plane exchange coupling on a topological surface is highly nontrivial
with respect to the case of out-of-plane exchange.
Indeed, in the case of an out-of-plane exchange a simple mean-field theory would generate a gap for arbitrarily small values of
the coupling constant $g$, and no semi-metal-insulator transition would take place in this case.\cite{Nogueira-Eremin-2012}
This situation is reminiscent from the metal-insulator transition in the Hubbard model, where a mean-field theory at
half-filling yields a gap $\Delta\sim e^{-{\rm const}/U}$, where $U$ is the on-site Coulomb interaction.\cite{Fradkin-Book}
It is well known that this result is not correct for values of $U$ smaller than energy scales
of the order of the bandwidth.\cite{Gebhard-book}
Note, that on a TI surface the mean-field result also leads to a phase transition if
a momentum dependence of the interaction induced by projecting the bulk Hamiltonian on the edge states is taken into
account but for some critical value $U_c$.\cite{Schmidt-2012}
In our case the out-of-plane exchange
is generated dynamically from the interplay between quantum planar magnetic and charge fluctuations,
which is characterized  by a competition between the in-plane exchange coupling $J$ and the screened Coulomb interaction $g$.
This leads to a dynamical mass generation accompanied by the generation of
a CS term, which implies a coupling between the in-plane magnetization and the electric field ${\bf E}=-\nablab a_0$, and
to a Berry phase governing the precession dynamics of the magnetization. Note that the inclusion of the charged channel in the
Hubbard-Stratonovich decoupling is crucial to obtain a CS term.

The case of in-plane magnetization is of great experimental relevance. Recently a thin film of FM insulator EuS has been
successfully grown on the surface of Bi$_2$Se$_3$,\cite{Moodera} making the surface of Bi$_2$Se$_3$ ferromagnetic
by proximity effect with magnetization at the interface being different from the bulk EuS values. There were several features which point out towards
a strong interaction between the magnetic moments of EuS and Bi$_2$Se$_3$. One of them refers to the fact that the dependence of the planar
magnetoresistivity of the interface Bi$_2$Se$_3$ shows an effectively lower Curie temperature than that of the bulk EuS which could be the result of the
quantum fluctuations due to presence of surface Dirac fermions.\cite{Nogueira-Eremin-2012} Another effect is even more interesting as it reports the significant
out-of-plane magnetization of the magnetic moments at the FM-TI interface while the bulk EuS has
the in-plane orientation of the magnetic moments.\cite{Moodera}
Our results show that the out of plane orientation of the
moments will be indeed generated at the interface by the interaction among
the Dirac fermions although on the experimental side further mechanisms related to
the crystalline anisotropy at the grown interface can be also in play. In addition, the direct gapping
of the Dirac spectrum of the surface electrons was reported very recently at the EuS-Bi$_2$Se$_3$ interface.\cite{kapitulnik-2013}
In particular, it was found that below Curie temperature there is a negative megnetoresistance near zero field which is believed
to be the consequence of gap opening in the Dirac spectrum due to proximity to the ferromagnet with out-of-plane magnetization.\cite{Lu-2011}
According to our calculations this effect must be dependent on the temperature and on the strength of the out-plane component
of the magnetization, induced by the interaction. This would be interesting to test experimentally.

\section{Conclusion}
\label{Sect:Conclusion}

In conclusion, we have shown that in topological insulators with a proximity-induced in-plane magnetization a gap
can be spontaneously generated by tuning the local electronic interaction above a critical value, leading in this way to
a semi-metal-insulator transition. In particular, considering the $N$ number of Dirac fermions we find that when $N$ is even, the masses are
generated in pairs $\pm|m|$ and no CS term is generated, so that parity and time-reversal is overall preserved.
On the other hand, for $N$ odd which the case of TI all generated masses are equal and positive and as a result,
the CS term with TME effect is generated. In particular, we find that the critical dimensionless value
of $\hat{g_c}$ for generating this term is also proportional to the value of the in-plane
exchange coupling of the FM making this effect to depend on the choice of the FM substrate in the experiment.

That no
CS term is generated for $N$ even is physically reasonable, since in this case we can change to a representation where
there are $N/2$ four-component Dirac spinors, in which case the model may be reinterpreted as some model for graphene,
a material featuring an even number of Dirac cones.
Interestingly, in such a graphene model the gap generation is
associated to a mass spectrum containing masses $\pm|m|$, a scenario not considered so far in interacting models for
graphene where the ``vector field'' has only the time component as compared to QED.\cite{interac-graphene,Sheehy-2007}
TIs, on the other hand, have an odd number of Dirac cones. In this context,
recent experiments on EuS/Bi$_2$Se$_3$ heterostructures open the possibility that the experimentally
elusive gap generation in QED-like theories may finally be observed in the near future.

\acknowledgments

We would like to thank Alex Altland, Igor Herbut, Jagadeesh Moodera, Achim Rosch, J\"org Schmalian, and Peng Wei
for interesting discussions. We thank J. S. Moodera for sending us his preprint prior publication.
The authors would like to thank the Deutsche Forschungsgemeinschaft (DFG) for the financial support via
the collaborative research center SFB TR 12.

\appendix
\section{Calculation of the vacuum polarization for massive two-component Dirac fermions coupled
to a gauge field}
\label{App:vacpol}

For pedagogical reasons, we review in this appendix in detail the calculation of the
one-loop vacuum polarization
in (2+1)-dimensional QED with massive two-component Dirac fermions.\cite{CS} We will perform the calculation
in Euclidean space (imaginary time). In this case the Dirac matrices satisfy the same algebra as the
Pauli matrices, having anticommutator  $\{\gamma_\mu,\gamma_\nu\}=\delta_{\mu\nu}$ and a commutator
$[\gamma_\mu,\gamma_\nu]=2i\epsilon_{\mu\nu\lambda}\gamma_\lambda$. From this algebra it follows the traces
of product of $\gamma$-matrices necessary to calculate the vacuum polarization,
\begin{equation}
\label{Eq:tr2gammas}
{\rm tr}(\gamma_\mu\gamma_\nu)=2\delta_{\mu\nu},
\end{equation}
\begin{equation}
 \label{Eq:tr3gammas}
 {\rm tr}(\gamma_\mu\gamma_\nu\gamma_\lambda)=2i\epsilon_{\mu\nu\lambda},
\end{equation}
\begin{equation}
 \label{Eq:tr4gammas}
 {\rm tr}(\gamma_\mu\gamma_\nu\gamma_\lambda\gamma_\rho)=2(\delta_{\mu\lambda}\delta_{\mu\rho}
 +\delta_{\mu\rho}\delta_{\lambda\nu}-\delta_{\mu\nu}\delta_{\lambda\rho}).
\end{equation}

\begin{figure}[h]
 \includegraphics[width=6cm]{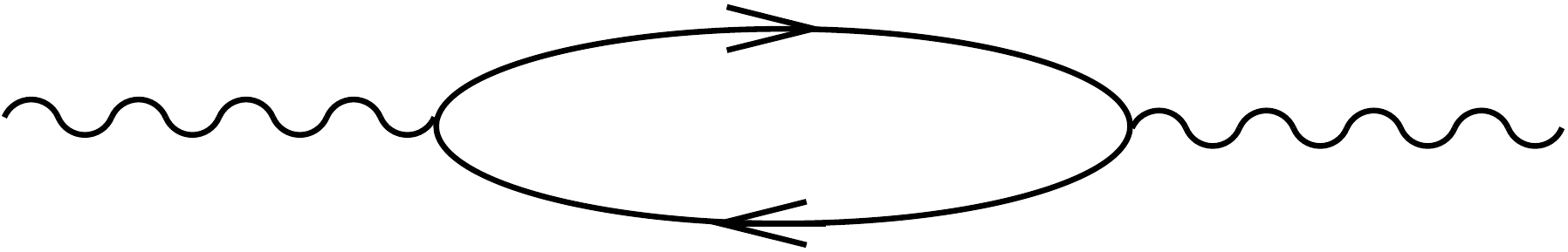}
 \caption{Feynman diagram representing the vacuum polarization. The wiggled lines represent photons and the internal
 lines represent fermionic propagators.}
 \label{Fig:vacpol}
\end{figure}

The vacuum polarization is represented by the Feynman diagram shown in Fig. \ref{Fig:vacpol}. It corresponds to
the photon self-energy and is analytically given by,
\begin{equation}
\label{Eq:QEDvacpol}
 \Sigma_{\mu\nu}(p)=-NJ^2\int\frac{d^3 k}{(2\pi)^3}{\rm tr}[\gamma_\mu G(k)\gamma_\nu G(p+k)],
\end{equation}
where $G(k)$ is the (matrix) fermion propagator,
\begin{equation}
 G(k)=\frac{1}{i\slashchar{k}+m}=\frac{m-i\slashchar{k}}{k^2+m^2},
\end{equation}
and the sign of the mass can be positive or negative. In the field theory literature $J=e$, the electric charge.

Using the trace formulas (\ref{Eq:tr2gammas},\ref{Eq:tr3gammas},\ref{Eq:tr4gammas}), we can express Eq. (\ref{Eq:QEDvacpol}) in
the form,
\begin{widetext}
\begin{equation}
\label{Eq:vacpol-1}
 \Sigma_{\mu\nu}(p)=2NJ^2\int\frac{d^3 k}{(2\pi)^3}
 \frac{2k_\mu k_\nu+k_\mu p_\nu+p_\mu k_\nu-\delta_{\mu\nu}(k^2+k\cdot p+m^2)}{(k^2+m^2)[(k+p)^2+m^2]}
 -2NJ^2m\epsilon_{\mu\nu\lambda}p_\lambda I(p),
\end{equation}
where
\begin{equation}
 \label{Eq:Integral}
 I(p)=\int\frac{d^3 k}{(2\pi)^3}\frac{1}{(k^2+m^2)[(k+p)^2+m^2]}.
\end{equation}
Current conservation implies that Eq. (\ref{Eq:vacpol-1}) can be cast in the form,
\begin{equation}
\label{Eq:tr-vacpol}
 \Sigma_{\mu\nu}(p)=2NJ^2S(p)\left(\delta_{\mu\nu}-\frac{p_\mu p_\nu}{p^2}\right)
 -2NJ^2m\epsilon_{\mu\nu\lambda}p_\lambda I(p).
\end{equation}
Indeed, it is not difficult to show that $p_\mu\Sigma_{\mu\nu}(p)=0$.
We will use dimensional regularization in this Appendix, with
the understanding that this is done only in the evaluation of integrals, while $\gamma$-matrices and Levi-Civitta tensor remain
as defined above for the (2+1)-dimensional case.

Taking the trace yields,
\begin{equation}
 \Sigma_{\mu\mu}(p)=4NJ^2 S(p)
 =-2NJ^2\int\frac{d^3 k}{(2\pi)^3}\frac{k^2+k\cdot p+3m^2}{(k^2+m^2)[(k+p)^2+m^2]}.
\end{equation}
By writing
\begin{equation}
 k^2+k\cdot p=\frac{1}{2}[(k^2+m^2)+(k+p)^2+m^2]-m^2-\frac{p^2}{2},
\end{equation}
we can express $S(p)$ in the form,
\begin{equation}
\label{Eq:S-integral}
 S(p)=-\frac{1}{2}\int\frac{d^3 k}{(2\pi)^3}\frac{1}{k^2+m^2}+\frac{1}{4}(p^2-4m^2)I(p).
\end{equation}
\end{widetext}
The rules of dimensional regularization imply,\cite{ZJ}
\begin{equation}
 -\int\frac{d^3 k}{(2\pi)^3}\frac{1}{k^2+m^2}=2m^2\int\frac{d^3 k}{(2\pi)^3}\frac{1}{(k^2+m^2)^2},
\end{equation}
such that Eq. (\ref{Eq:S-integral}) becomes,
\begin{equation}
 S(p)=p ^2\Pi(p)=m^2I(0)+\frac{1}{4}(p^2-4m^2)I(p),
\end{equation}
where we have introduced the standard notation $\Pi(p)$ used in QED.

The integral $I(p)$ can be evaluated explicitly,
\begin{equation}
 I(p)=\frac{1}{4\pi|p|}\arctan\left(\frac{|p|}{2|m|}\right)=\frac{1}{4\pi|p|}\arcsin\left(\frac{|p|}{\sqrt{p^2+4m^2}}\right),
\end{equation}
and
\begin{equation}
 I(0)=\frac{1}{8\pi|m|}.
\end{equation}
We note therefore the important limit cases,
\begin{equation}
 \lim_{m\to 0} S(p)=\frac{|p|}{32},
\end{equation}
and
\begin{equation}
 S(p)=\frac{p^2}{48\pi |m|}+{\cal O}(p^4).
\end{equation}
Therefore, in the small mass limit the photon self-energy becomes,
\begin{equation}
 \Sigma_{\mu\nu}(p)\underset{m\ll|p|}{\approx}\frac{NJ^2|p|}{16}\left(\delta_{\mu\nu}-\frac{p_\mu p_\nu}{p^2}\right)
 -\frac{NJ^2m}{4|p|}\epsilon_{\mu\nu\lambda}p_\lambda,
\end{equation}
which in terms of the one-loop vacuum polarization for massless fermions
as defined in the main text, $\Pi(p)=NJ^2/(16|p|)$ becomes,
\begin{equation}
\label{Eq:small-mass}
 \Sigma_{\mu\nu}(p)\underset{m\ll |p|}{\approx} p^2\Pi(p)\left(\delta_{\mu\nu}-\frac{p_\mu p_\nu}{p^2}\right)
 -4m\Pi(p)\epsilon_{\mu\nu\lambda}p_\lambda.
\end{equation}
In the large mass limit, on the other hand, we obtain,
\begin{equation}
\label{Eq:large-mass}
 \Sigma_{\mu\nu}(p)\underset{|p|\ll m}{\approx}\frac{NJ^2}{24\pi m}p^2\left(\delta_{\mu\nu}-\frac{p_\mu p_\nu}{p^2}\right)
 -\frac{NJ^2}{4\pi}\frac{m}{|m|}\epsilon_{\mu\nu\lambda}p_\lambda.
\end{equation}
The above form was used in Ref. \onlinecite{Nogueira-Eremin-2012} to derive the effective
Lagrangian for the magnetization dynamics on the surface of a three-dimensional TI.

\section{Vector field propagator}
\label{App:vecprop}

In order to find the propagator (\ref{Eq:gauge-prop}) of the vector field $a_\mu$ from Eq. (\ref{Eq:S-eff}), we
have simply to solve the matrix equation,
\begin{equation}
\label{Eq:matrixeq}
 M_{\mu\alpha}(p)D_{\alpha\nu}(p)=\delta_{\mu\nu},
\end{equation}
where
\begin{equation}
\label{Eq:Mmunu}
 M_{\mu\nu}(p)=\Pi(p)(\delta_{\mu\nu}p^2-p_\mu p_\nu)+L_\mu L_\nu,
\end{equation}
with $L_\mu=(J/\sqrt{g})\delta_{\mu0}$.
In order to solve Eq. (\ref{Eq:matrixeq}), we
decompose $D_{\mu\nu}(p)$ in the form,
\begin{equation}
 D_{\mu\nu}(p)=A\delta_{\mu\nu}+Bp_\mu p_\nu+CL_\mu L_\nu+Dp_\mu L_\nu+Ep_\nu L_\mu.
\end{equation}
The unknown coefficients $A$, $B$, $C$, $D$, and $E$ are easily determined
from Eq. (\ref{Eq:matrixeq}), so that  Eq. (\ref{Eq:gauge-prop}) follows.

The calculation can be easily generalized to the case where a CS term is present and Eq. (\ref{Eq:Mmunu}) is
replaced by
\begin{equation}
\label{Eq:Mmunu-1}
 M_{\mu\nu}(p)=\Pi(p)(\delta_{\mu\nu}p^2-p_\mu p_\nu)+L_\mu L_\nu+\frac{4}{N}\sum_im_i\Pi(p)\epsilon_{\mu\nu\lambda}p_\lambda,
\end{equation}
where we have assumed that the masses are small (see Appendix \ref{App:vacpol}).
In this case the decomposition of $D_{\mu\nu}(p)$ has to include the additional tensors
$\epsilon_{\mu\nu\lambda}p_\lambda$, $\epsilon_{\mu\alpha\lambda}p_\lambda L_\alpha p_\nu$, and
$\epsilon_{\mu\alpha\lambda}p_\lambda L_\alpha p_\mu$. The result for the propagator is thus,
\begin{eqnarray}
 D_{\mu\nu}(p)&=&\frac{1}{(p^2+16m^2)\Pi(p)}\left\{\delta_{\mu\nu}+\frac{[(p^2+16m^2)\Pi(p)+L^2]}{(p\cdot L)^2}
 p_\mu p_\nu\right.\nonumber\\
 &-&\left.\frac{p_\mu L_\nu+p_\nu L_\mu}{p\cdot L}-\frac{4}{Np^2}\sum_i m_i\epsilon_{\mu\nu\lambda}p_\lambda\right\},
\end{eqnarray}
which in the regime $|p|\gg |m|$ used to solve the gap equation and
to write Eq. (\ref{Eq:Sodd}) is approximated simply by,
\begin{eqnarray}
 D_{\mu\nu}(p)&\approx&\frac{1}{p^2\Pi(p)}\left\{\delta_{\mu\nu}+\frac{[p^2\Pi(p)+L^2]}{(p\cdot L)^2}
 p_\mu p_\nu\right.\nonumber\\
 &-&\left.\frac{p_\mu L_\nu+p_\nu L_\mu}{p\cdot L}-\frac{4}{Np^2}\sum_i m_i\epsilon_{\mu\nu\lambda}p_\lambda\right\}.
\end{eqnarray}

\section{Evaluation of integrals}
\label{App:ints}

Here we calculate two integrals appearing in Eq. (\ref{Eq:self}):
\begin{equation}
 I=\int\frac{d^3k}{(2\pi)^3}\frac{|k|}{\omega^2(k^2+m^2)},
\end{equation}
and
\begin{equation}
 J=\int\frac{d^3k}{(2\pi)^3}\frac{k^2}{\omega^2(k^2+m^2)},
\end{equation}
where we recall the notation $k=(\omega,v_F{\bf k})$ and $k^2=\omega^2+v_F^2{\bf k}^2$. Here we can simply set $v_F=1$.
The integrand of both integrals $I$ and $J$ are singular at $\omega=0$. However, we can solve this singularity by assuming a
regularization where a finite value is obtained through the principal values of $I$ and $J$. Let us consider first the integral $I$.
Using partial integration, we can cast the integral in $\omega$ in the form,
\begin{widetext}
\begin{equation}
 \int_{-\infty}^\infty\frac{d\omega}{2\pi}\frac{\sqrt{\omega^2+{\bf k}^2}}{\omega^2(\omega^2+{\bf k}^2+m^2)}
 =\int_{-\infty}^\infty\frac{d\omega}{2\pi}\frac{1}{\omega}\frac{d}{d\omega}\left(\frac{\sqrt{\omega^2+{\bf k}^2}}{\omega^2+{\bf k}^2+m^2}
 \right)=\int_{-\infty}^\infty\frac{d\omega}{2\pi}\frac{1}{\sqrt{\omega^2+{\bf k}^2}}\left[\frac{2m^2}{(\omega^2+{\bf k}^2+m^2)^2}
 -\frac{1}{\omega^2+{\bf k}^2+m^2}\right].
\end{equation}
\end{widetext}
Thus, we can write
\begin{equation}
 I=2m^2\int\frac{d^3k}{(2\pi)^3}\frac{1}{|k|(k^2+m^2)^2}-\int\frac{d^3k}{(2\pi)^3}\frac{1}{|k|(k^2+m^2)}.
\end{equation}
For $\Lambda\gg |m|$, we have,
\begin{equation}
 I=\frac{1}{2\pi^2}\left[1+\ln\left(\frac{|m|}{\Lambda}\right)\right].
\end{equation}

The integral $J$ can be rewritten as $J=J_1+J_2$, where
\begin{equation}
 J_1=\int\frac{d^3k}{(2\pi)^3}\frac{1}{k^2+m^2},
\end{equation}
and
\begin{equation}
J_2= \int\frac{d^3k}{(2\pi)^3}\frac{{\bf k}^2}{\omega^2(k^2+m^2)}.
\end{equation}
In the limit $\Lambda\gg |m|$ we can trivially evaluate $J_1$:
\begin{equation}
 J_1=\frac{\Lambda}{2\pi^2}-\frac{|m|}{4\pi}.
\end{equation}
For $J_2$ we can again regularize the singularity for $\omega=0$,
\begin{eqnarray}
 J_2&=&-\int\frac{d^2k}{(2\pi)^2}\int_{-\infty}^\infty\frac{d\omega}{2\pi}\frac{{\bf k}^2}{\omega^2+{\bf k}^2+m^2}\frac{d}{d\omega}\frac{1}{\omega}
 \nonumber\\
 &=&2\int\frac{d^2k}{(2\pi)^2}\int_{-\infty}^\infty\frac{d\omega}{2\pi}\frac{{\bf k}^2}{(\omega^2+{\bf k}^2+m^2)^2}
 \nonumber\\
 &=&\frac{1}{4\pi}\int_0^\Lambda dk\frac{k^3}{(k^2+m^2)^{3/2}}=\frac{1}{4\pi}(\Lambda-2|m|).
\end{eqnarray}
Therefore,
\begin{equation}
 J=\frac{\Lambda(1+\pi)}{4\pi^2}-\frac{3|m|}{4\pi}.
\end{equation}

\end{document}